\documentclass[aps,preprint,epsfig,rotate,showpacs]{revtex4}
\usepackage{epsfig}
\usepackage{graphicx}
\usepackage{amsmath}
\def\hatt{\hat{T}}
\def\fqqzero{\hat{F}_Q(\hat{q},0)}

\newcommand{\beq}{\begin{eqnarray}}
\newcommand{\eeq}{\end{eqnarray}}
\begin{document}

\title{Chaos and Correspondence in  Classical and Quantum
Hamiltonian Ratchets: A Heisenberg Approach}
\author{Jordan Pelc$^{1}$, Jiangbin Gong$^{2,3}$ and Paul Brumer$^{1}$}
\affiliation{$^{1}$Chemical Physics Theory Group, Department of
Chemistry, and Centre for Quantum Information and Quantum Control,
University of Toronto, Toronto, M5S 3H6, Canada\\
$^{2}$Department of Physics and Centre of Computational Science and
Engineering, National University of Singapore, 117542, Singapore\\
$^{3}$ NUS Graduate School for Integrative Sciences and Engineering,
117597, Singapore}
\date{\today}

\vspace{1.0in}

\begin{abstract}

Previous work [Gong and Brumer, Phys. Rev. Lett., 97,
240602 (2006)] motivates this study as to how
asymmetry-driven quantum ratchet
effects can persist despite a corresponding fully chaotic classical
phase space. A simple perspective of ratchet dynamics, based on the
Heisenberg picture, is introduced. We show that ratchet effects are
in principle of common origin in classical and quantum mechanics,
though full chaos suppresses these effects in the former but not
necessarily the latter. The relationship between ratchet effects and
coherent dynamical control is noted.
\end{abstract}

 \maketitle

\section{Introduction}

Originally proposed by Smoluchowski \cite{Smoluchowsky} and Feynman
\cite{Feynman}, and motivated by an application to biological
molecular motors \cite{bio}, studies of ratchet transport, that is,
asymmetry-driven directed transport without external bias, are now
the subject of an expanded range of theoretical interest
\cite{Reimann,Hanggitoday}.  While earlier investigations depended
on external noise to rationalize these directional effects, recent
work has shown that they can persist even in its absence
\cite{Flach,Flach2,gonghanggi,Schanz,gongpre04,lundh,Harper,Poletti,kenfack},
thereby raising questions about the origin of transport in
isolated Hamiltonian systems. Many studies have therefore focused on the
relationship between ratchet dynamics and deterministic chaos (see,
for instance, \cite{Schanz,gongpre04,Harper}),
relating ratchet transport to the typical questions of chaology,
including, naturally, the complex relationship between quantum
systems and their corresponding chaotic classical counterparts. In
the context of recent cold-atom testing of Hamiltonian ratchet
transport in classically chaotic systems
\cite{test,sadgrove2007,dana2008}, investigations of ratchet
transport are interesting both as a method of exploring quantum and
classical transport properties as well as a means of addressing general
questions in quantum and classical chaos.

It has been shown \cite{lundh,Harper,Poletti} that quantum ratchet
transport is possible even when the corresponding classical dynamics
is completely chaotic.  In such a case, the classical system displays
no appreciable current. Hence, these systems show a novel
qualitative divergence between quantum and classical dynamical
properties, motivating this study of the relationship between
quantum and classical ratchet transport.

Below we show that ratchet
effects emerge, both quantum mechanically and classically, via an
asymmetry-induced distortion of the spatial distribution, leading to a net
effective force.  Classically, full chaos diminishes this
distortion, and hence suppresses ratchet effects. Quantum mechanically, by
contrast, the distortion generally persists, except at very small
values of the effective Planck constant.

Hamiltonian ratchet dynamics is also directly related to
laser-induced coherent control of directional transport
\cite{book,Ignacio}. Symmetry-breaking schemes have been used in
coherent control since its inception \cite{photocurrent}, and
so studying quantum vs. classical ratchet transport also lends
insight into quantum control scenarios. Recent results
\cite{Ignacio} suggest that such control, once thought to be
exclusively quantum mechanical, is possible classically, as well.
Quantum ratchet transport in the presence of full classical chaos,
by contrast, is an excellent example of controlled transport that
may not be possible in classical dynamics. As such, this topic
is also of interest to two related, more general issues: quantum
controllability of classically chaotic systems; and survival
conditions for quantum control in the classical limit.

The case of quantum ratchet transport with full classical chaos
discussed below further strengthens the view that quantum control of
classically chaotic systems is often feasible
\cite{gongprl2001,chaosreview}. Indeed, this interesting possibility
has already attracted some interest, both theoretically and
experimentally \cite{fujisaki2007,sadgroveEPL,fujisaki2008}.

We consider here spatially-periodic quantum systems with Hamiltonian
$\hat{H}=\hat{H}_0(\hat{p}) + \hat{V}(\hat{q},t)$, where
$\hat{V}(\hat{q},t)$ is a time-periodic operator representing an
external potential imposed on the system, $\hat{q}$ and $\hat{p}$
are conjugate position and momentum operators, respectively, and
operators are denoted by a circumflex.  These systems display
ratchet transport, that is, despite being initially distributed
uniformly in space, having zero initial momentum, and being driven
by a force without bias, they organize to show an increase in the
current or the average momentum, denoted $\langle p \rangle$ below
for both quantum and classical mechanics.  The absence of a biased
force means that upon averaging over all space, denoted by an
overbar:
\begin{eqnarray}
\overline{-\frac{\partial V(q,t)}{
\partial q}} \equiv \overline{ F(q,t)} = 0,
\end{eqnarray} where $V(q,t)$ and $F(q,t)$ are the coordinate-space
representation of the applied potential and force, respectively
\cite{rocking-ratchet-note}. Significantly, this zero average, which
is the standard definition of the absence of bias, is entirely
independent of the structure or state of the physical system upon
which $F(q,t)$ acts. As such, as emphasized below,
$\overline{F(q,t)}$ is conceptually distinct from the expectation
value of a net force $\langle F(t) \rangle$ actually felt by an
evolving system, which, of course, is a function of the system
evolution. The significance of this distinction will become apparent
in what follows.

\begin{figure}
\epsfig{file=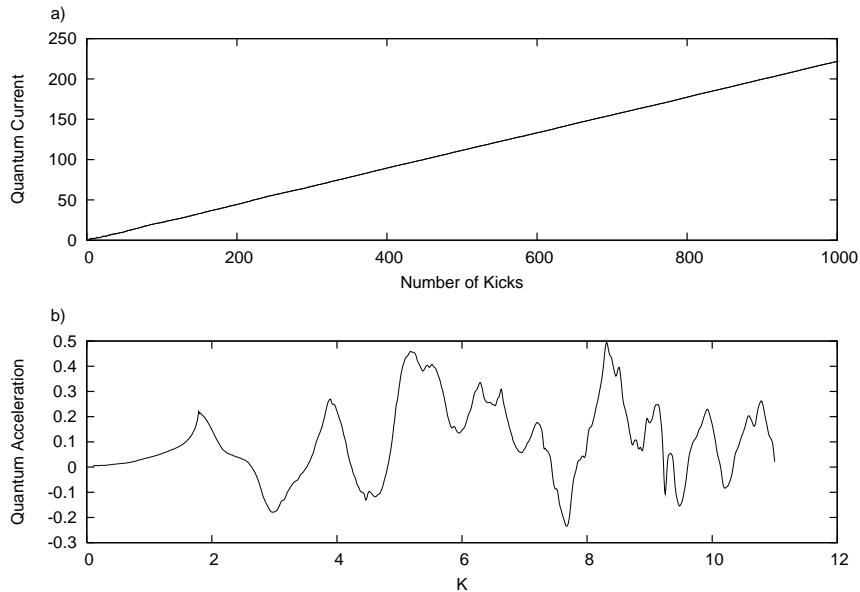,width=11.5cm} \caption {(a) Time dependence
of the  quantum current $\langle p \rangle_Q$ of the modified kicked
Harper system for $K = 4$, $J = 2$, and $\hbar = 1$, shown here for
the first 1000 kicks, with the initial state given by a momentum
eigenstate with zero momentum.  (b) The mean acceleration rate of
the quantum current for a range of $K$ values, with $K=2J$ and
$\hbar=1$. Note that, as seen in Ref. [28], the transport direction
may change erratically with the initial condition. As explained in
the text, all quantities here and those in other figures are in
dimensionless units.}
\end{figure}

While the discussion presented below is quite general, we continue
to employ our modification of the kicked Harper
paradigm \cite{Harper,Sheppy,wanggong} as an illustrative example
and motivator of this study. The quantum modified Harper Hamiltonian
is given by \cite{scale-note}
\begin{eqnarray}
\label{Hami}
\hat{H} & = & J\cos(\hat{p})+K\hat{V}_r(\hat{q})\sum_{n}\delta (t-n); \\
\hat{V}_r(\hat{q})& = & \cos(2\pi \hat{q}) +
\sin(4\pi \hat{q}), \label{Pot}
\end{eqnarray}
where the system potential is $\hat{V}(\hat{q},t) =
K\hat{V}_r(\hat{q})\sum_{n}\delta (t-n)$, and we define
$\hat{F}_r(\hat{q}) \equiv -\frac{\partial \hat{V}_r(\hat{q})}
{\partial \hat{q}}$. Here $t$ is the time, $n$ is an integer, and
$J$ and $K$ are system parameters.  The associated $q$-space is
periodic in $[0,1]$.   All system variables here should be
understood to be appropriately scaled and hence dimensionless. In
particular, the scaled, dimensionless Planck constant is denoted as
$\hbar$, and hence $\hat{p}=-i\hbar \frac{\partial}{\partial
\hat{q}}$. The unitary time evolution operator associated with one
kick from time $t=0$ to $t=1+\epsilon$
in Eq. (\ref{Hami}) is given by
\begin{eqnarray}
\hat{U}(1,0) & = & e^{-\frac{iJ}{\hbar}\cos(\hat{p})}
e^{-\frac{iK}{\hbar}\hat{V}_r(\hat{q})}, \label{Prop}
\end{eqnarray}
with the cumulative time evolution operator from $t=0$ to $t=m$ given by
\begin{equation}
\hat{U}(m,0) = [\hat{U}(1,0)]^m.
\end{equation}
We also stress that the initial quantum state used here is always
assumed to be a zero momentum eigenstate, which is time-reversal
symmetric and spatially uniform. As shown in Ref. [28], the ratchet
transport can be a sensitive function of the initial state. However,
our analyzes below can be easily adapted to other initial states.

The properties of this model discussed below hold in general for the
regime where the kicked Harper does not show dynamical localization
\cite{localnote}. We consider the case where $K = 2J$, although this
choice is arbitrary. Since this system can be exactly mapped onto
the problem of a kicked charge in a magnetic field \cite{Dana}, or
can be related to cold-atom experiments \cite{Coldatom,wanggong} or
to driven electrons on the Fermi surface \cite{Fishman}, it has a
realistic physical and experimental interpretation.  In particular,
though the unkicked part of the Hamiltonian in Eq. (2) is given by
$J\cos(\hat{p})$, the underlying dispersion relation in the
cold-atom and kicked-charge realizations of the kicked Harper model
is still given by $E=\hat{p}^2/2$ \cite{Dana,Coldatom,wanggong}.
That is, the momentum variable in our abstract model can still be
directly linked to the mechanical momentum of a moving particle.
Hence the current of particles can indeed be calculated via the
momentum expectation value.

The quantum dynamics associated with the propagator in Eq.
(\ref{Prop}) shows unbounded \cite{unbounded} acceleration of the
ratchet current \cite{Harper}.
 Typical results are
shown in Fig. 1, where panel (a) shows the current $\langle p
\rangle$ for $K=4$ and panel (b) shows the mean current acceleration
rate as a function of $K$. Here the acceleration is defined
approximately as $\langle p(t=1000) \rangle/1000$.

The classical comparison with the quantum dynamics
considers ensembles of trajectories that
are analogous to the quantum systems discussed above: initially,
the trajectories have zero momentum and are uniformly distributed
in coordinate space, and are driven by a force of zero spatial
mean at all times.  Again, our discussion is quite general for such
systems, although we consider the classical analogue of the modified
kicked Harper system as an illustrative example, obtained by
replacing the quantum operators in Eqs. (\ref{Hami}) and (\ref{Pot})
with their respective classical
observables.
Specifically, the evolution of a classical
trajectory through one kick is then given by
\begin{eqnarray}
\label{PProp}
p_N & = & p_{N-1} + KF_r(q_{N-1}) \nonumber \\
q_N & = & q_{N-1} - J\sin(p_N).
\label{QProp}
\end{eqnarray}
This system has been shown to display
virtually no classical ratchet transport \cite{Harper} if the
system parameters are in the regime of full classical chaos.

Throughout this discussion, it will often be convenient to consider
quantum and classical arguments simultaneously.  We distinguish
quantum and classical objects by respective subscripts $Q$ and $C$, and
refer to both dynamics when these subscripts are omitted.

This paper is organized as follows.  Section II analyzes, from a
new perspective based on the Heisenberg picture of the dynamics, the
origin of asymmetry-driven ratchet transport.  Section III considers
the difference and correspondence between classical and
quantum ratchet transport. Section IV summarizes the conclusions of
this study.

\section{Asymmetry and Ratchet Effects}

\subsection{The Heisenberg Force}

Classically and quantum mechanically, the rate of the ratchet current increase,
here termed the acceleration, at time $t$ is given by the expectation
value of the net force at that time:
\begin{eqnarray}
{\frac{d \langle p \rangle }{dt}} & = & \langle F(t) \rangle. \label{Current}
\end{eqnarray}
Evidently, $\langle p\rangle$ must remain zero if it begins at zero
and $\langle F(t)\rangle=0$ at all times. Hence, when ratchet
acceleration occurs, it follows that the expectation value of the
net force must be nonzero.  This result calls for analysis of how
ratchet acceleration is possible in the absence of a biased force.

To facilitate comparison of quantum and classical
mechanics, it is convenient to cast this discussion in
terms of the density matrix formalism.
The expectation value of the quantum force at time $t$ is given by
\begin{eqnarray}
\langle F(t) \rangle_Q & = {\rm Tr} \left[\hat{\rho}_Q(t) \fqqzero\right]= &
{\rm Tr} \left[\hat{\rho}_Q(0)\hatt e^{-\frac{i}{\hbar}\int_0^t dt'\hat{L}_Q(t')}
\fqqzero\right],
\label{ensemble}
\end{eqnarray}
where $\hat{\rho}_Q(0)$ is the (pure state) density matrix at time
zero and $\hat{\rho}_Q(t)$ is the propagated density at time $t$
\cite{notation}. Here, time evolution is mediated by the quantum
Liouville operator
$\hat{L}_Q\cdot=\frac{i}{\hbar}[\hat{H},\cdot]$, the bracket $[$ ,
$]$ is the commutator, and $\hatt$ denotes the time-ordering
operator. For the Hamiltonian  in  Eq. (\ref{Hami}), $\fqqzero =
-K
\partial V_r(\hat{q})/\partial \hat{q}$.

The effect of the time-ordered exponential is given in terms of
the evolution operator as \cite{Zubarev}
\begin{equation}
\hat{F}_{Q,H}(\hat{q},t=n)= \hat{U}^{-1}(n,0) \fqqzero
\hat{U}(n,0)
\end{equation}
Equation (\ref{ensemble}) can be rewritten as \cite{Zubarev}
\begin{eqnarray}
\langle F(t) \rangle_Q & = & {\rm Tr} \left[\hat{\rho}_Q(t)
\fqqzero\right] = {\rm Tr} \left[\hat{\rho}_Q(0)
\hat{F}_{Q,H}(\hat{q},t) \right], \label{shorter}
\end{eqnarray}
where
\begin{equation}
\hat{F}_{Q,H}(\hat{q},t) \equiv \hatt e^{-\frac{i}{\hbar}
\int_0^t\hat{L}_Q(t')dt'}\hat{F}_Q(\hat{q},0)
\end{equation}
defines the {\em Heisenberg force}, the focus of attention below.

The classical, ensemble-averaged value of the force at time $t$ is
similarly given by
\begin{eqnarray}
\langle F(t) \rangle_C & = & \int
dp dq \left[\rho_{C}(0)
 \hatt e^{-i\int_0^t dt' \hat{L}_C(t') } F_C(q,0)
\right],
\label{class-eq}
\end{eqnarray}
where $\rho_C(0)$ is the initial classical density distribution,
$\hat{L}_{C}\cdot=i\{H,\cdot\}$ is the classical Liouville
operator, where $\{$  , $\}$ represents a classical Poisson
bracket, and $F_C(q,0) = -K \partial V_r(q)/\partial q$. The time
evolution of $q$, and hence of $F_C(q,0)$, is carried out via
Eq. (\ref{QProp}).

For either the quantum or the classical ensemble average $\langle
F(t) \rangle$ to be nonzero, and hence induce ratchet
acceleration, some system  attribute  needs to break the
positive-negative symmetry to ``choose'' a direction.  From Eqs.
(\ref{ensemble}) and (\ref{class-eq}) it is clear that asymmetries
in either the initial distribution, force, or evolution operator
are essentially equivalent as the origin of bias. Since, for
classical and quantum ratchets, the initial distribution and force
are chosen to be symmetric, the asymmetry in the evolution
operator, and hence asymmetry in dynamics induced by the
Hamiltonian, must be responsible for the nonzero net current.

Specifically, consider the $q$-representation of Eq. (\ref{shorter}).
Noting that
$\hat{\rho}_Q(0)$ describes a spatially uniform state (i.e.
$\hat{\rho}_Q(0) = |q \rangle\langle q| )$, in
normalized coordinates
$\rho_Q(q,0) \equiv \langle q| \hat{\rho}_Q(0)|q\rangle = 1$,
so that
\begin{eqnarray}
\langle F(t) \rangle_Q & = &
\int dq \langle q|\hat{\rho}_Q(t)|q \rangle \langle q | \fqqzero |q \rangle =
\int dq \langle q | \hat{\rho}_Q(0) |q \rangle \langle q| \hat{F}_{Q,H} (\hat{q},t)|q \rangle \\
&=& \int dq \langle q |\hat{F}_{Q,H} (\hat{q},t)|q \rangle =
\overline{\hat{F}_{Q,H} (\hat{q},t)} \label{forceform}
\end{eqnarray}
That is, the average force is dictated by  the uniform spatial
average over the Heisenberg force, as distinguished from the {\it
Schr\"odinger force} $\hat{F}_Q(\hat{q},0)= -K \partial
\hat{V}(\hat{q})/\partial \hat{q}$. Correspondingly, since
$\rho_C(0)$ is chosen to be normalized and spatially uniform, Eq.
(\ref{class-eq}) indicates that
\begin{eqnarray}
\langle F(t) \rangle_C = \overline{\hatt e^{-i\int_0^t dt' \hat{L}_C(t')
}F_C(q,0)} = \overline{F_{C,H}(q,t)} , \label{classform}
\end{eqnarray}
i.e.,  a spatial average over the time-evolving classical force
$F_{C,H}(q,t)$, analogous to the quantum case. Since Eqs.
(\ref{forceform}) and (\ref{classform}) show that the expectation
value of the force is given by an average over the {\it evolving}
force, a nonzero net force as a result of an asymmetry in the
dynamics becomes possible, even if the spatial average of the bare
force $F(q,t)$ itself remains zero at all times.

Note that, since the force is diagonal in $q$ in quantum
mechanics, and not a function of $p$ classically, the evolving
force distribution $F_H(q,t)$ is adequately described entirely in
$q$ in both mechanics.  This allows simple, direct comparisons of
quantum and classical mechanics, as shown in the following
section. Below, we term $F_{Q,H}(q,t) \equiv \langle q |
\hat{F}_{Q,H}(\hat{q},t)| q \rangle$ the force distribution in $q$
associated with the Heisenberg force. Similar terminology applies
in classical mechanics.  The diagonal element of the
$q$-representation of the distribution of the Schr\"odinger
density $\langle q | \hat{\rho}_Q(t)|q \rangle$, is denoted
$\rho_Q(q,t)$, so that $\langle F(t) \rangle_Q = \int dq
\rho_Q(q,t)F_Q(q,0)$. The classical object analogous to
$\rho_Q(q,t)$ is the $q$-component of the evolving density,
$\rho_C(q,t) \equiv \int dp \rho_C(p,q,t)$, where $\rho_C(p,q,t)$
is the classical evolving density.  In both mechanics, the initial
spatial distribution is assumed uniform. As a result, in the
quantum case for example, and in accord with Eqs. (\ref{ensemble})
and ({\ref{shorter}),
\begin{eqnarray}
F_{Q,H}(q,t) &=& \langle q|\hat{F}_{Q,H}(\hat{q},t)|q \rangle =
\langle q| \hat{\rho}_Q(0) \hat{F}_Q(\hat{q},t)| q \rangle \nonumber
\\
& =&  \langle q| \hat{\rho}_Q(t)|q \rangle \langle q|\hat{F}_Q(\hat{q},0)|q
\rangle = \rho_Q(q,t)F_Q(q,0)
\end{eqnarray}
That is, the evolving force
distribution is given by the bare force weighted by the evolving
density. The analogous result holds in classical mechanics.

Given that, in either mechanics, $F_H(q,t) = \rho(q,t)F(q,0)$,
with a uniform initial distribution $\rho(0)$ and unbiased force
$F(q,0)$, a net nonzero $F_H(q,t)$  requires that $\rho(q,t)$
weights $F(q,0)$ so as to break the directional symmetry.
Minimally, the system evolution must be such that each point $q_i$
in the $q$-space does not in general have a complement $q_j$ such
that both $\rho(q_i,t)=\rho(q_j,t)$ and $F(q_i,0)=-F(q_j,0)$. This
is the simplest asymmetry condition on the dynamics necessary for
the generation of a ratchet current. The modified Harper
Hamiltonian [Eq. (\ref{Hami})] clearly satisfies this condition.

This Heisenberg approach thus gives a simple picture of ratchet
current generation. The origin of a current arising from a net
force can be understood as either as (a) a distortion in the
density $\rho(q,t)$, which will weight the bare force $F(q,0)$
non-uniformly giving rise to a nonzero average, or (b) as a
distortion in the evolving force $F_{H}(q,t)$ itself, whose
average $\langle F(t) \rangle$ is nonzero due to this distortion,
even if the bare force has zero mean. The advantage of using the
evolving force picture is that it resolves the intuitive puzzle of
how directional transport in the momentum space emerges in the
absence of a biased force. Whether the force itself, that is, the
bare force, is biased or not is irrelevant. Rather, the intrinsic
asymmetry in the dynamics permits the evolving force $F_H(q,t)$ to
develop a nonzero mean, and hence a nonzero ratchet acceleration
rate.

\begin{figure}
\begin{center}
\epsfig{file=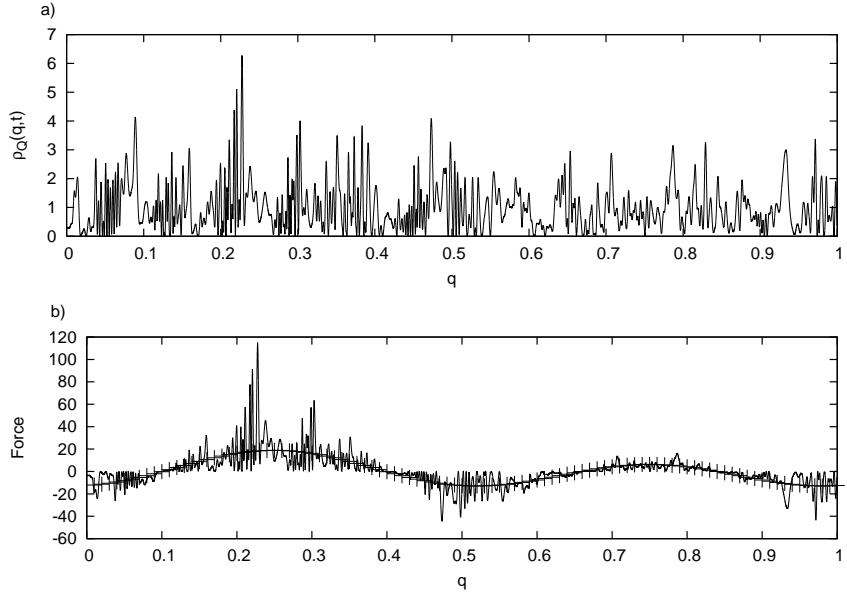,width=11.5cm}
\end{center}
\caption {(a) $\rho_Q(q,t)$ and (b) $F_{Q,H}(q,t)$ compared to
$F_Q(q,t)$ (plus symbols) for the modified kicked Harper model after
the first 50 kicks for $K=4$, $J=2$ and $\hbar=1$. Distortion in the
density $\rho_Q(q,t)$, and hence bias in the Heisenberg force
distribution function $F_{Q,H}(q,t)$, are evident.}
\end{figure}

Computationally, the Heisenberg picture is easily applied to
the modified kicked Harper model to examine $F_{Q,H}(q,t)$.  For example,
Fig. 2 shows $\rho_Q(q,t)$ and $F_{Q,H}(q,t)$ for parameters
associated with an appreciable and unbounded ratchet current
acceleration.  Despite starting with a flat distribution in $q$,
$\rho_Q(q,t)$ in Fig. 2(a) is now clearly unevenly distributed.  Accordingly,
the distribution of the Heisenberg force $F_{Q,H}(q,t)$ shown in
Fig. 2(b) is strongly biased compared to the symmetric bare force
distribution (plus symbols).

\subsection{Two Roles of the Force}

Implicitly, we have considered the force in two capacities: acting
on the structure of the ensemble and thereby producing a nonzero net
Heisenberg force; and the net force itself, acting within an
ensemble average to generate ratchet acceleration, i.e. $\langle
F(t) \rangle = {\frac{d \langle p \rangle }{dt}}. $ To further
elucidate how this relates to ratchet transport, consider any
$\delta$-kicked quantum ratchet model with an arbitrary kicking
potential operator $K \hat{V}_r(\hat{q})$ and kinetic energy
operator $J\hat{T}(\hat{p})$. The evolution of this type of system
is mediated by a propagator $\hat{U}$ like Eq. (\ref{Prop}), such
that a Heisenberg observable $\hat{O}_{Q,H}$ after $N$ kicks is
given by $\hat{O}_{Q,H}(N) = (\hat{U}^{-1})^N \hat{O} \hat{U}^N$.
Consider the current $\langle p(1)\rangle_Q$ after the first kick:
\begin{eqnarray}
\langle p(1) \rangle_Q & = &
{\rm Tr}[\hat{\rho}_Q(0) \hat{U}^{-1} \hat{p} \hat{U} ] \nonumber \\
& = & {\rm Tr}[\hat{\rho}_Q(0) e^{ \frac {iK}{\hbar} \hat{V}_r(\hat{q})}
e^{ \frac{iJ}{\hbar}\hat{T}(\hat{p})} \hat{p}
 e^{ \frac{-iJ}{\hbar}\hat{T}(\hat{p})} e^{ \frac {-iK}{\hbar}
\hat{V}_r(\hat{q})} ].
\end{eqnarray}
Using $\hat{p}=-i\hbar\frac{\partial}{\partial q}$ and that the
initial state is assumed uniform in $q$, one obtains
\begin{eqnarray}
\langle p(1) \rangle_Q & = & -K {\rm Tr}[\hat{\rho}_Q(0)  e^{ \frac {iK}{\hbar}
\hat{V}_r(\hat{q})} e^{ \frac {-iK}{\hbar} \hat{V}_r(\hat{q})} \frac{\partial
\hat{V}_r(\hat{q})}{\partial \hat{q}}] \nonumber \\
&&  + {\rm Tr}[\hat{\rho}_Q(0) \hat{p}] \nonumber \\
& = & -K\overline {\frac{\partial \hat{V}_r(\hat{q})}
{\partial \hat{q}}}+ 0 = 0.
\label{P1}
\end{eqnarray}
This
illustrates the distinction between the force's role in distorting its own
distribution and its role in inducing a current.
That is, although no current develops after the first kick, subsequent
kicks produce current.  Therefore, even though the net force remains
zero for the first kick, that kick distorts the
system so that it will subsequently experience a net force.

More generally, for $N$ kicks,
\begin{eqnarray}
\langle p(N) \rangle_Q & = &
{\rm Tr}[\hat{\rho}_Q(0) (\hat{U}^{-1})^N \hat{p} \hat{U}^N ]\nonumber \\
& = & {\rm Tr}[\hat{\rho}_Q(0)  (e^{ \frac {iK}{\hbar}
\hat{V}_r({\hat{q}})}e^{ \frac{iJ}{\hbar}\hat{T}(\hat{p})})^N \hat{p}
(e^{ \frac {-iK}{\hbar} \hat{V}_r({\hat{q}})}e^{
\frac{-iJ}{\hbar}\hat{T}(\hat{p})})^N ] \nonumber \\
& = & -K \sum_{j=0}^{N-1} {\rm Tr}[\hat{\rho}_Q(0)
(\hat{U}^{-1})^j \frac{\partial \hat{V}_r({\hat{q}})}{\partial {\hat{q}}}
\hat{U}^j ] \nonumber \\
& = & K\sum_{j=0}^{N-1}\langle F_r(j) \rangle_Q.
\label{PN}
\end{eqnarray}
It follows that the change in $\langle p \rangle $ on each step is
\begin{eqnarray}
\Delta \langle p \rangle_Q \equiv \langle p(N) \rangle_Q  - \langle p (N-1)
\rangle_Q  = K \langle F_r(N-1) \rangle_Q,
\label{DeltaP}
\end{eqnarray}
showing that the change in momentum induced at every kick is a
result of the net force from the previous kick.
This makes clear
the general case:  the force first acts on an ensemble to generate a
distortion, and then a net ratchet force can develop. Exactly the
same arguments apply in classical mechanics.


\begin{figure}
\epsfig{file=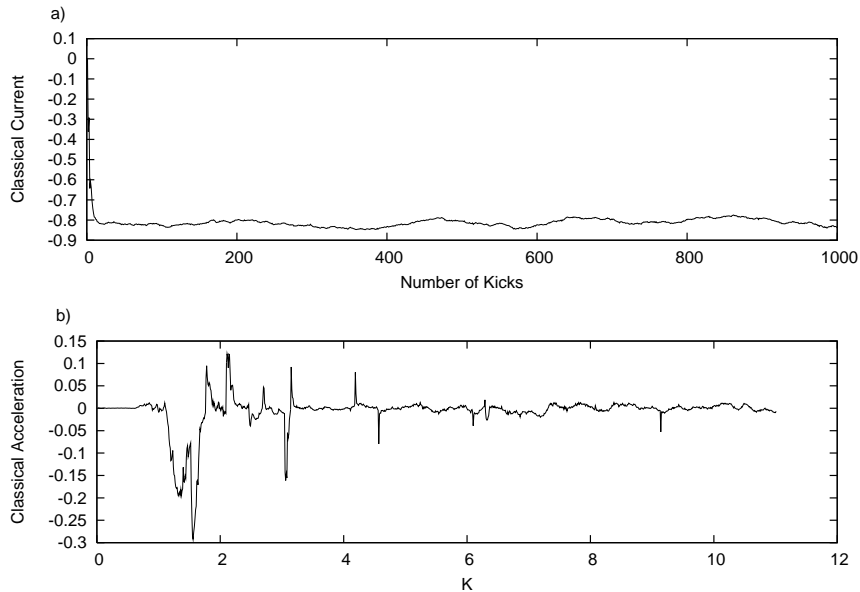,width=11.5cm} \caption {(a) Time
dependence of the classical current $\langle p \rangle_C$ of the
modified kicked Harper system for $K = 3$ and $J = 1.5$, shown here
for the first 1000 kicks.   (b)  The mean acceleration rate of the
classical current for a range of $K$ values, with $K=2J$.}
\end{figure}

Thus far, this discussion has supported the view \cite{Ignacio} that
symmetry-breaking induced transport can be achieved both classically
and quantum mechanically, both arising via a distortion originating
from an asymmetry in the dynamics. However, despite the existence of
this analogous ratchet transport mechanism in the classical modified
Harper model, classical ratchet transport behaves very differently
in the regime of classically chaotic motion, where the classical
current quickly saturates at a value close to zero. This is clear in
Fig. 3: panel (a) shows the saturating current $\langle p \rangle_C$
for a typical chaotic case; and panel (b) shows the classical mean
acceleration rates for a range of parameters. When $K$ is greater
than approximately 3.7, the classical dynamics develops full chaos and
the mean acceleration rate is generally negligible.
(The occasional isolated nonzero mean acceleration rates seen above
$K \approx$ 3.7
are likely due to some remnants of pre-chaotic structure in
phase space). Therefore, even
though the relevant symmetry properties are the same classically and
quantum mechanically, some other important distinction must exist.

\section{Evolution of Classical and Quantum Hamiltonian Ratchets}

Results in Fig. 3 demonstrate that the behavior of the classical
modified kicked Harper system is quite different from the quantum
result, where unbounded ratchet effects persist.  If indeed ratchet
effects emerge by the same mechanism in quantum and classical
mechanics, it remains to be explained why that mechanism generates
different results for different mechanics.  Specifically, ratchet effects
diminish classically in the regime associated with classical chaos.
We therefore examine how the
onset of chaos affects classical ratchet dynamics, in a way that does not occur
quantum mechanically.  We also discuss the peculiar long-time behavior of the quantum
modified kicked Harper model, as well as quantum-classical correspondence.

\subsection{Chaos and the Heisenberg Force}

From a trajectory perspective, classical chaos is characterized by
exponential sensitivity to initial conditions.  However, the
conventional interpretation of quantum mechanics does not
describe individual trajectories. Hence,
a comparison of quantum and classical dynamics demands comparison of
quantum and classical distributions
\cite{Rice,Gu,Ballentine2,Ballentine,Pattanayak2,Pattanayak,Wilkie,GBLyapunov}.
Although KAM theory and
finite-time limitations suggest deviations in the properties of
distribution functions of typical classically chaotic systems from theoretical
ideals \cite{Zaslavsky}, such systems are still
expected to exponentially develop increasingly fine structure.
Upon coarse graining on the scale of interest, the classical phase space
distribution in a fully chaotic system uniformly fills the phase
space almost everywhere,
with additional structure detectable only on an increasingly fine
scale. Indeed, this is what is termed full chaos in most numerical
studies of this type: when no structure is visible in the phase
space on a pre-set fine scale, it is considered operationally chaotic.

Consider then how this applies to the Heisenberg force for the ratchet
systems considered here, where the phase space is always
bound or periodic in $q$.  Ensemble
averages are computed by integrating over the distribution.  Since complete
chaos implies no structure in $\rho_C(q,t)$
on the scale of interest, such averages will
look like unweighted averages in $q$
(provided that the scale on which the
variable of interest varies is much larger than the scale of
structure in $\rho_C(q)$ remaining in the chaotic phase space).  That is, we
can essentially ignore the $q$-component of the density when taking
spatial averages. In the case of the force,
\begin{eqnarray}
\langle F(t) \rangle_{C}  & = & \int dp dq \rho_C(p,q,t)
F_C(q,t) \nonumber \\
& \approx & \int dq F_C(q,t)
\int dp \rho_C(p,t) \nonumber \\
& \propto & \int dq F_C(q,t) = 0, \label{ergo2}
\end{eqnarray}
where $\rho_C(p,t)=\int dq \rho_C(p,q,t)$ is the classical momentum
density distribution. Hence, for all times when the phase space is
operationally chaotic, the ensemble average of the classical force is
proportional to the spatial average of the bare force: i.e., zero. Chaotic
dynamics here implies no spatial distortion of the system on the
scale of interest, and hence no creation of a net evolving force.
This is consistent with a result of the
``classical sum rule" \cite{Schanz}, which predicts
that there will be no classical ratchet current in fully chaotic
systems.



The comparison with quantum mechanics is straightforward.  If
the quantum $q$-distribution $\rho_Q(q,t)$ is flat,
or if the scale of structure remaining
in this distribution is far smaller than that over which the bare force
$F_Q(q,0)$ varies, then by an argument analogous to the classically chaotic
case, the net quantum force $\langle F(t)\rangle_Q$ will be essentially zero.
As in the classical case, the spatial distortion giving rise to a net
force would not be appreciable on the scale of interest.

However, a quantum ratchet system is not expected to display such
behavior. The Fourier relationship between $\rho_Q(q,t)$ and
$\rho_Q(p,t) \equiv \langle p | \hat{\rho}_Q(t) | p \rangle$
implies that a uniform distribution in space $\rho_Q(q,t)=1$
corresponds to the lowest momentum state
$\rho_Q(p,t)=\delta_{p,0}$.  Once the system is driven by a force,
other momentum states will of course be populated.
Correspondingly, $\rho_Q(q,t) = \sum_{k,k'} c_k c^*_{k'}
e^{-\frac{i}{\hbar}(p_k-p_{k'})q}$, where the $c_k$ are constants
and $p_k$ are momenta. This density is not flat.  For fixed $p_k$
and $p_{k'}$, a sufficiently large $\hbar$ can always be found so
that the $e^{-\frac{i}{\hbar}(p_k-p_{k'})q}$ terms oscillate
sufficiently slowly, giving $\rho_Q(q,t)$ structure in $q$ on the
scale of interest. Therefore, sufficiently far into the quantum
regime, driven quantum systems are expected to retain coarse
structure in $q$-space; there is a limit to the fineness of scale
in quantum mechanics \cite{Eckhardt}. Consequently, the net force
is not in general expected to reduce to the average bare force.

This provides a qualitative explanation
for the difference in behavior between quantum and
classical dynamics in the regime of full classical chaos.
This perspective also accounts for the difference in controllability
between classical and quantum mechanics.  That is, asymmetry-driven transport
control is in principle possible in both.  Since it relies on a
distortion of the system distribution function, distributions
without structure on the relevant scale show diminished control.
Classically, control is therefore lost to chaos, whereas it can survive
in quantum mechanics.

\subsection{Quantum Long-Time Dynamics}

To achieve stable, unbounded acceleration of the ratchet current,
as observed in the modified Harper system, requires that $\langle
F(t) \rangle_Q$ continually operate in the same direction, driving
a current with essentially the same bias for all time. This
implies that the profile of the time-evolving density
$\rho_Q(q,t)$, and hence of the Heisenberg force distribution
$F_{Q,H}(q,t)$, does not change appreciably in time (or that it
changes in the highly unlikely way that always maintains the same
bias).  If the quasienergy spectrum of the system is purely
discrete, this can not be the case. Specifically, from Floquet
theory we have that for any time-periodic, bounded quantum system
with discrete quasienergy spectrum, the density is given by
$\rho_{Q}(q,t) = \sum_{l,l'} d_l d^*_{l'}
e^{\frac{i}{\hbar}(E_l-E_{l'})t} \rho_{Q}(q,0)$, where the $d_l$
are constants and the $E_l$ are the quasienergies \cite{Hogg}.
Since this density is the sum of periodic functions, it is itself
quasiperiodic.  Therefore, ensemble averages in such systems are
also quasiperiodic, and hence do not continuously increase in time
\cite{Hogg,Peres}. This is true as well for the the Heisenberg
force $ F_{Q,H}(q,t)$, which would be quasiperiodic and hence
eventually reverse its direction.

For this reason, earlier quantum ratchet models without current
saturation occurred for kicked-rotor systems with quantum
resonance conditions  \cite{lundh,resonance}, displaying a
continuous quasi-energy spectrum. The behavior of the modified
kicked Harper model here, which apparently does not satisfy a
quantum resonance condition, and for which extended computational
results (not shown here) have suggested unbounded directional
current, therefore requires explanation.

In fact, it can be shown that the all kicked Harper systems can be exactly
mapped onto the problem of a kicked charge in a magnetic field, although
{\it only at resonance} \cite{Dana}.  Consequently, the quasienergy spectrum
of this model is not necessarily purely discrete, the system evolution need not
be quasiperiodic and the modified kicked
Harper system need not necessarily show dynamical saturation in time.

\begin{figure}
\epsfig{file=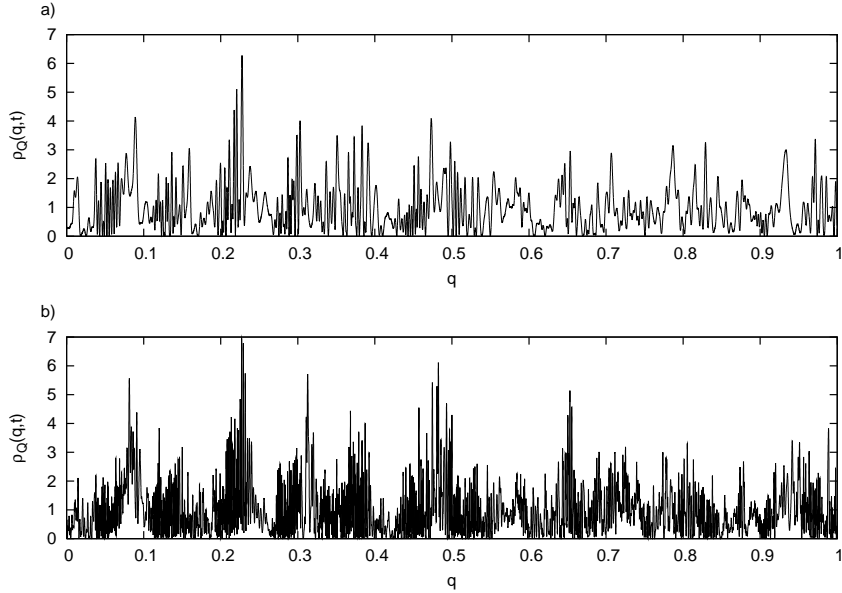,width=11.5cm} \caption {$\rho_Q(q,t)$ of the
modified kicked Harper system for $K=4$, $J=2$, and $\hbar=1$, after
(a) the first 50 kicks,  and (b) the first 200 kicks. Note that the
probability distribution function in (b) oscillates more drastically
than in (a), but their overall shape remains roughly the same. This
is consistent throughout the parameter space.}
\end{figure}

As an example, Fig. 4 shows $\rho_Q(q,t)$ after $50$
and $200$ kicks for typical parameters, and Fig. 5 shows
$F_{Q,H}(q,t)$ compared to the bare force $F_{Q}(q,0)$ for the same
circumstances. Indeed, there is no appreciable change in the
qualitative shape of either $\rho_Q(q,t)$ and $F_{Q,H}(q,t)$ after the
first few kicks, although the very fine details of the oscillatory
structure increase.



\begin{figure}
\epsfig{file=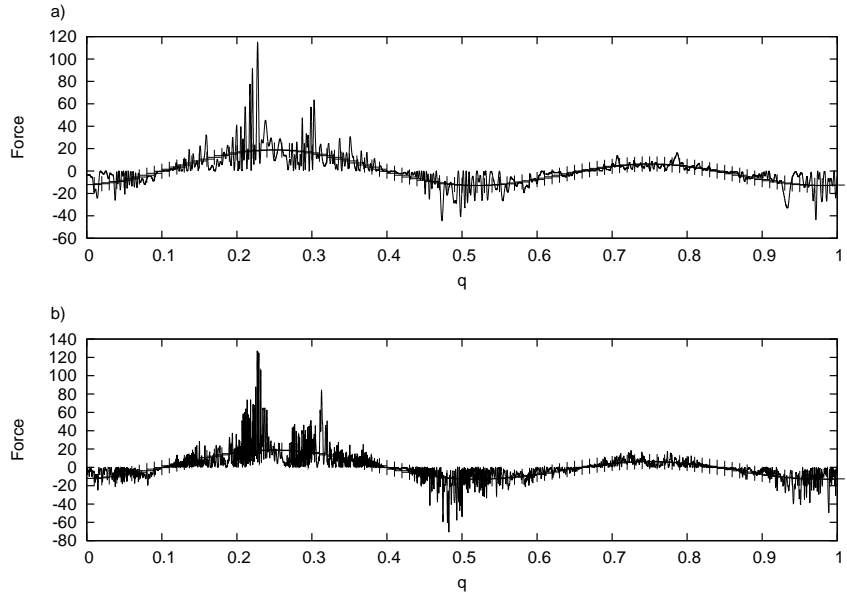,width=11.5cm} \caption {$F_{Q,H}(q,t)$
of the modified kicked Harper system compared to $F_{Q}(q,t)$ (plus
symbols) for $K=4$, $J=2$, and $\hbar=1$, after (a) the first 50
kicks,  and (b) the first 200 kicks. Note that the Heisenberg force
distribution function in (b) oscillates more drastically than in
(a), but their overall shape remains roughly the same. This is
consistent throughout the parameter space.}
\end{figure}




\subsection{Quantum-Classical Correspondence}

Given the above-mentioned quantum-classical differences, it is
natural to ask how the classical results emerge from the
quantum mechanics as the effective Planck constant $\hbar$ decreases.

Before resorting to computational studies, let us first examine how the
quantum dynamics may appear more classical for small $\hbar$. Consider
a time-evolving quantum density $\rho_Q(q,t) = \sum_{k,k'} c_k
c^*_{k'} e^{-\frac{i}{\hbar}(p_k-p_{k'})q}$, where the $c_k$ are
constants and $p_k$ are momenta. For large $\hbar$ the interference
between different momentum components induces large-scale patterns
in the density. However, for sufficiently small $\hbar$ relative to
$(p_k -p_{k'})q$, the exponential factor
will rapidly oscillate; the smallest scale of structure can
be much finer than the scale over which the bare force changes.
Hence, at a given time, and for smaller and smaller $\hbar$,
the quantum limit on the fineness of scale diminishes.  As in classical
mechanics, coarse scale structure can persist, but it no longer
has to.  Therefore, it
becomes possible for the ensemble-averaged quantum force to either maintain
an appreciable bias, as in the classically partially-integrable regime,
or to approach its average over a flat distribution, as in the classically
chaotic regime.  Qualitatively, then, the
coarse-scale structure in $q$ imposed by quantum coherence
can diminish as $\hbar \rightarrow 0$.

\begin{figure}
\epsfig{file=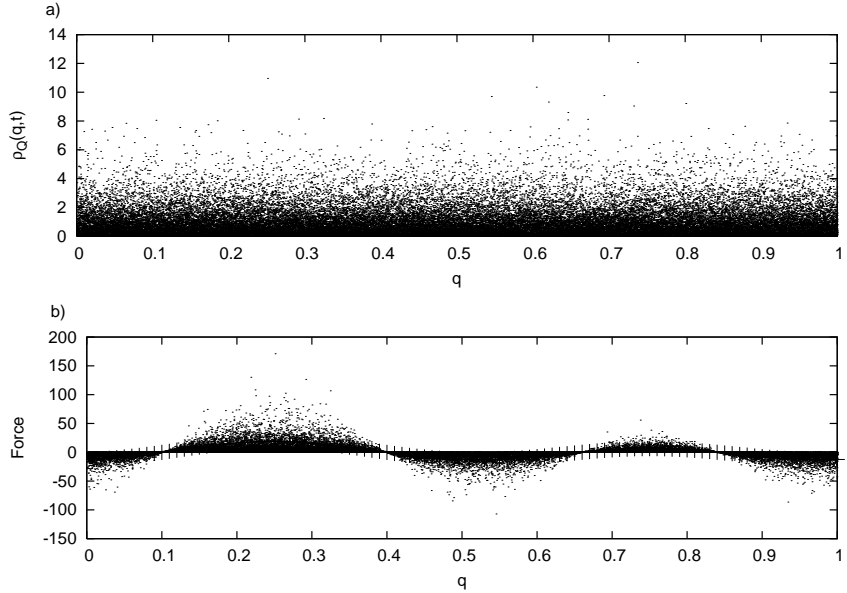,width=11.5cm} \caption {(a)
$\rho_Q(q,t)$ and (b) $F_{Q,H}(q,t)$ compared to $F_Q(q,t)$ (plus
symbols) for the modified kicked Harper model after the first 50
kicks for $K=4$, $J=2$ and $\hbar=0.0001$. }
\end{figure}

Figure 6 shows the $q$-representation of $\rho_Q(q,t)$,
as well as a comparison of the quantum Heisenberg and
Schr\"{o}dinger force distributions, $F_{Q,H}(q,t)$ and
$F_{Q}(q,0)$, for a typical case in a semiclassical regime,
represented by $\hbar=0.0001$ (a computationally-intensive regime).  The
system parameters here are associated with classical chaos.
The density in Fig. 6(a)  shows clear, truly drastic, oscillations,
with a roughly uniform oscillation amplitude. Further, it is evident
from Fig. 6(b) that on this scale, the overall distribution of the
Heisenberg force is similar to that of the initial Schr\"odinger
force distribution, justifying the loss of directional effects in
going from quantum to classical mechanics.
Figure 7 shows the quantum ratchet current $\langle p \rangle_Q$ in
the semiclassical regime of $\hbar=0.0001$, as compared with the
corresponding classical current $\langle p \rangle_C$. The quantum
current $\langle p \rangle_Q$ remains close to zero, and mimics the
classical current $\langle p \rangle_C$ almost exactly.

\begin{figure}
\epsfig{file=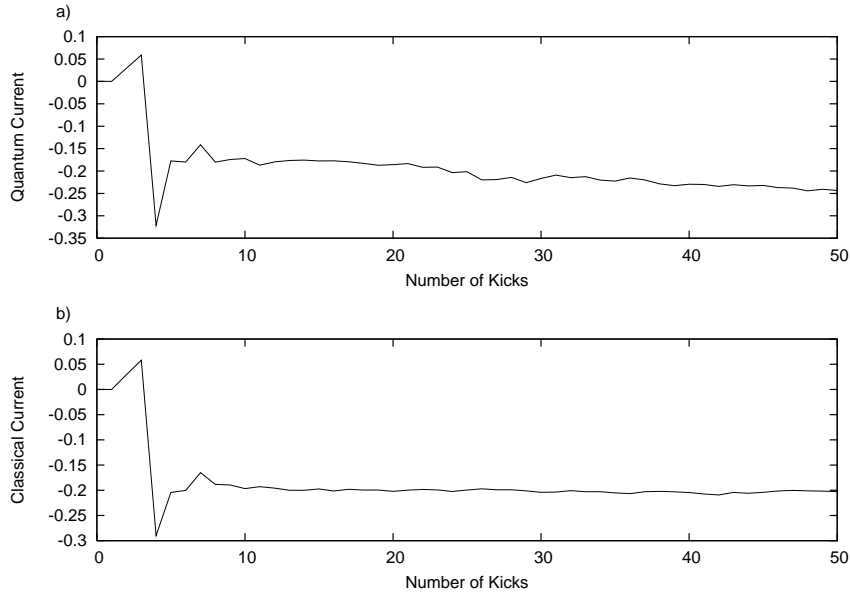,width=11.5cm} \caption{Comparison of the
(a) $\hbar=0.0001$ quantum current $\langle p \rangle_Q$ with (b)
its classical analogue $\langle p \rangle_C$ in the modified kicked
Harper system for the first 50 kicks for $K=4$ and $J=2$.}
\end{figure}

\section{Conclusion}

We sought here to explain certain general features of  ratchet
transport in Hamiltonian systems, and in particular to explain the
quantum vs classical behavior of the ratchet accelerator model
developed in Ref. \cite{Harper}.

Here we have introduced, and applied, the concept of a Heisenberg
or evolving force, in both quantum and classical mechanics, to
ratchet transport.  This showed that whether the bare force (i.e.,
the external force applied to the system) is unbiased is
irrelevant, since it is the evolving force that actually affects
net transport.  In both mechanics, asymmetry in the dynamical
evolution can cause asymmetric spatial distortion which leads to
the development of a net force and a nonzero current.
Symmetry-breaking-based control of quantum and classical transport
is hence of the same origin.


However, quantum and classical ratchet systems behave differently due to
chaos. Classical systems fail to generate ratchet current when their
phase space is fully chaotic, as the system distortion is
effectively canceled, and the asymmetry that leads to
directionality is lost. A completely chaotic phase space forces
ensemble averages to reduce to phase-space means that are
independent of the detailed aspects of the dynamics. In such cases
the ensemble-averaged net force remains zero for a non-biased
external force. By contrast, the equivalent effect is
prevented in quantum mechanics, where coarse-scale structure is preserved.
Symmetry-breaking-based quantum
control of transport in classically chaotic systems is hence possible. For the
same reason, quantum ratchet transport with full classical chaos
becomes a strong indication of non-chaotic properties of the quantum
dynamics.

The peculiar feature of the modified kicked Harper system, that it shows
unbounded linear transport for a wide parameter regime, is explained by its
mapping onto a resonant system, and hence having a continuous spectrum.  Its
dynamics therefore is not necessarily quasiperiodic.
Further, we computationally showed that if the quantum
system is sufficiently close to the classical limit, then quantum
ratchet behavior smoothly approaches classical ratchet behavior.

The advantage of using the Heisenberg force to gain insight into
the ratchet dynamics is expected to be generalizable to other
systems.


{\bf Acknowledgments:} J.P. and P.B. are supported by a grant from the
National Sciences and Engineering Research Council of Canada. J.G.
is supported by the start-up fund, (WBS No. R-144-050-193-101 and
No. R-144-050-193-133) and the NUS ``YIA" fund (WBS No.
R-144-000-195-123), both from the National University of Singapore.


\begin{thebibliography}{100}
\bibitem{rocking-ratchet-note}
A rocking ratchet constitutes a different case, where the absence of
a biased force means that the average of the force is zero when
averaging over an interval of time.
\bibitem{scale-note}
This model was originally \cite{Harper} proposed with
$\hat{V}_r(\hat{q})= \cos(\hat{q})+\sin(2\hat{q})$.
The spatial period has been rescaled for convenience.
\bibitem{unbounded} Throughout this paper we use the term unbounded
acceleration to denote the results of computational studies carried
out for long times (and indeed longer than those shown in this
paper). However, there is no formal proof that such dynamics shows
unbounded acceleration for times longer than those that were
computed.
\end{thebibliography}
\end{document}